\documentclass[namedreferences]{solarphysics}
\usepackage{natbib}                     	% For citations: redefine \cite commands
\usepackage[optionalrh]{spr-sola-addons} 	% For Solar Physics
\usepackage{graphicx}
\usepackage{color}                       	% For color text: \color command
\usepackage{url}                         	% For breaking URLs easily trough lines
\newcommand{\etal}{ et al. }
\newcommand{\be}{\begin{equation}}
\newcommand{\ee}{\end{equation}}
\newcommand{\beq}{\begin{eqnarray}}
\newcommand{\eeq}{\end{eqnarray}}

\newcommand{\aap}{    {\it Astron. Astrophys.}}

\newcommand{\apj}{    {\it Astrophys. J.}}
\newcommand{\apjl}{   {\it Astrophys. J. Lett.}}
\newcommand{\apss}{   {\it Astrophys. Spa. Sci.}}

\newcommand{\grl}{    {\it Geophys. Res. Lett.}}

\newcommand{\jgr}{    {\it J. Geophys. Res.}}
\newcommand{\mnras}{  {\it Mon. Not. Roy. Astron. Soc.}}
\newcommand{\nat}{    {\it Nature}}

\newcommand{\solphys}{{\it Solar Phys.}}

\newcommand{\rd}{\color{red}} %KLK
\begin{document}
\begin{article}
\begin{opening}
\title{Wavelet Analysis on Solar Wind Parameters and Geomagnetic Indices}
\author{Ch. \surname{Katsavrias}$^{1}$,
	P. \surname{Preka-Papadema}$^{1}$,
	X. \surname{Moussas}$^{1}$}
%-------------------------------------------------
\runningauthor{Katsavrias \etal}
\runningtitle{Wavelet Analysis on Solar Wind Parameters and Geomagnetic Indices}
\institute{$^{1}$ Department of Physics of Astrophysics, Astronomy and Mechanics, Department of Physics, 
			University of Athens, Panepistimiopolis Zografos (Athens) , GR-15783, Greece
		email: \url{ckatsavrias@phys.uoa.gr}
		email: \url{ppreka@phys.uoa.gr}  
		email: \url{xmoussas@phys.uoa.gr} \\}
%-------------------------------------------------
\begin{abstract} 
{The sun as an oscillator produces frequencies which propagate in the heliosphere, via solar wind, to
the terrestrial magnetosphere. We searched for those frequencies in the parameters of the near Earth solar 
plasma and the geomagnetic indices for the past four solar cycles. The solar wind parameters used in this work are 
the interplanetary magnetic field, plasma beta, Alfven Mach number, solar wind speed, plasma temperature, 
plasma pressure, plasma density and the geomagnetic indices $D_{ST}$, AE, $A_p$ and $K_p$. We found out that 
each parameter of the solar wind exhibit certain periodicities which differentiate in each cycle. 
Our results indicate intermittent periodicities in our data, some of them shared between the solar wind parameters 
and geomagnetic indices.}
\end{abstract}
%-------------------------------------------------
\end{opening}
%-------------------------------------------------
\section{Introduction}\label{Introduction}

{\rd Solar activity is quantified, mostly, by sunspot (or Wolf) number and exhibits a basic 11 years 
periodicity with 5.5 and 3.7 years harmonics \citep{Currie1976,Carta1982,Djurovic1996,Mursula1997,Polygiannakis2003}).} 
The quasi--biennial oscillation and the 1.3 years periodicity, 
short--term periodicities of 154, 128, 102, 77, 51 and 27 days have 
also been observed in the number of sunspots as well as in solar energetic 
events such as flares \citep{Kilic2009} and coronal mass ejections \citep{Lou2003}. 
{\rd The identification, interpretation and modelling of solar periodicities is an open 
research issue:}

\begin{itemize}

\item The 11 years periodicity in the mumber of sunspots and other solar 
activity \citep{Djurovic1996,Carta1982,Mursula1997}, is believed to originate from 
magnetohydrodynamic dynamo action at the base of the convection zone.
 
\item Based on magnetic helicity observations, \citet{Bao1998}, proposed a double-cycle
solar magnetic dynamo model with one dynamo operating at the base of the convection zone (11-year cycle) 
due to radial shear and the other at the top of the convection zone (quasi--biennial periodicity) due 
to latitudal shear \citep{Benevolenskaya1998,Benevolenskaya2000}.

\item It has also been proposed that the sunspot number can be well fitted by a superposition of the usual 11-yr cycles and wave trains 
with periodicity continuously varying from 3 years at solar maximum, to 1.7 years towards solar minimum \citep{Polygiannakis2003}.

\item An 154-day periodicity was observed in the solar flare rate, the 10.7 cm radio flux, sunspot number and the global magnetic 
field of the Sun \citep{Gonzalez1993,Lean1989,Cane1998}. \citet{Bai1993},  interpret this periodicity (51, 77, 102 and 128 days as well) 
as sub harmonics of a solar clock mechanism with a fundamental of 25.5 days. On the other hand, \citet{Rieger1984} proposed the g-mode 
oscillations as a possible cause while \citet{Wolff1992} a combination of two r-modes with an interior 
g-mode \citep[see also~][for a discussion on Rossby wave origin of the Rieger periodicities]{Dimitropoulou2008,Dimitropoulou2009}. 

\item The 1.3 years oscillation was found in solar magnetic field during the maxima and declining phase of 8 solar cycles. 
This is thought to be generated at the tachocline as an inherent feature of the global magnetic field \citep{Obridko2007,Howe2000}. 
The same periodicity appears in sunspot number and other solar parameters; \citet{Krivova2002}
interprete it, as a sub harmonic, of the 154 days period mentioned above.
 
\item A wavelet analysis by \citet{Polygiannakis2002}, in the interval 1994 -- 2000, indicates ephemeral periodicities of 27, 59, 137 
and 330 days for flares and 37, 97, 182 and 365 days for CMEs. 
\end{itemize}

Solar wind, being the continuation of solar corona, exhibits also periodic behaviour possibly be driven by the Sun periodic 
variations. The terrestrial magnetic field-solar wind interaction triggers magnetospheric activity (ionospheric electric currents, auroras, 
magnetic storms, turbulence and magnetic reconnection) quantified by the global magnetospheric indices. As the interaction depends on the 
solar wind physical characteristics, the indices are expected to exhibit similar periodicities as the solar wind. Therefore, the study of 
the periodicities of the solar wind and the geomagnetic indices lead to understanding the connection between them and the solar activity. 

This work examines these joint solar wind and magnetospheric variations within the 1966--2010 interval which includes four solar cycles. 
The time series examined exhibit a non stationary, quasi-periodic behaviour where periodic components appear intermittently and with 
varying significance levels. We use, therefore, wavelet analysis in our study in addition to the classic Lomb/Scargle 
periodogram \citep{Lomb1976,Scargle1982} which is used for comparison.
%-------------------------------------------------------------------------------------
\begin{figure}[ht]
\centering 
\resizebox{\textwidth}{!}{\includegraphics{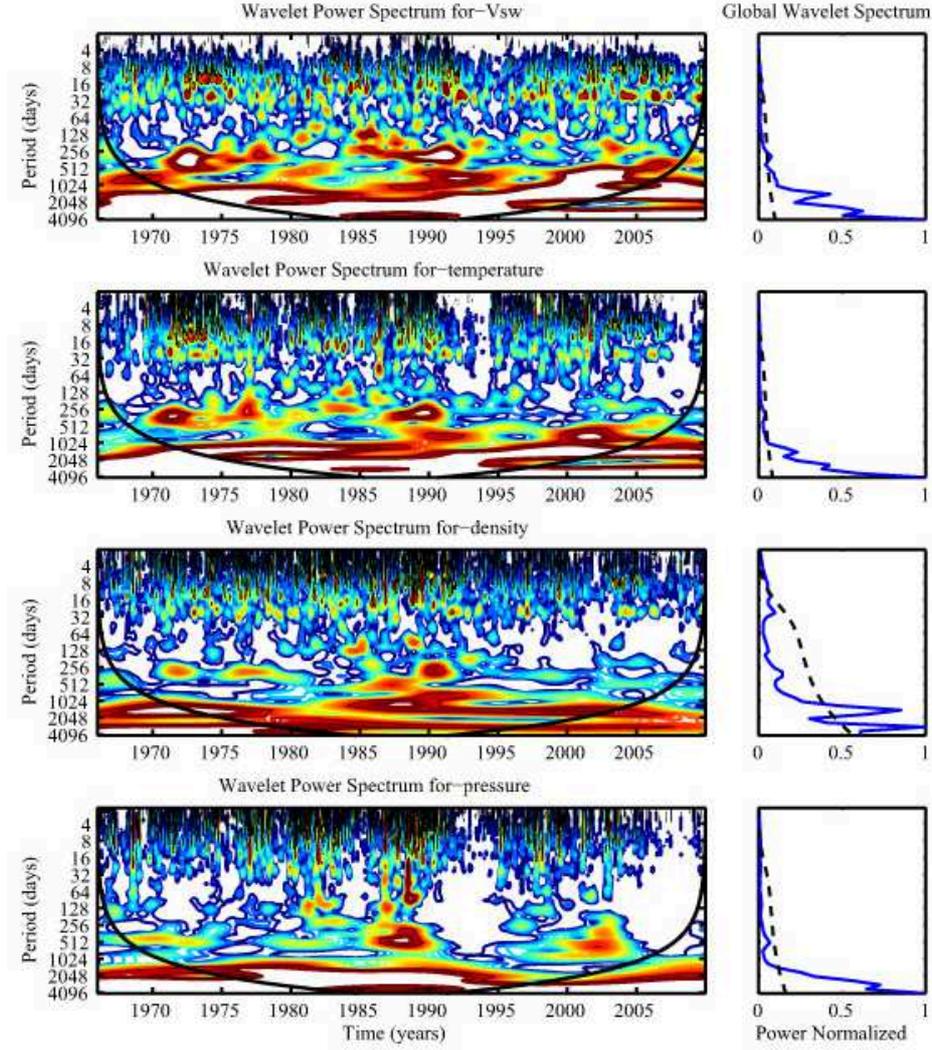}}
\caption{Wavelet spectra of solar wind parameters (Top to bottom): speed ($V_{sw}$), plasma temperature (T), 
plasma density (D) and plasma pressure (P). (Our plot is colour--coded  with red corresponding to the 
maxima; the black contour is the 99 \% confidence level). The normalized global wavelet spectra {\rd (see \ref{WAnal})} are shown on the right
of each panel{\rd ; in this case the dashed lines represent confidence level above 99 \%}.).}
\label{F1}
\end{figure}
%-------------------------------------------------------------------------------------
\begin{figure}[ht]
\centering 
\resizebox{\textwidth}{!}{\includegraphics{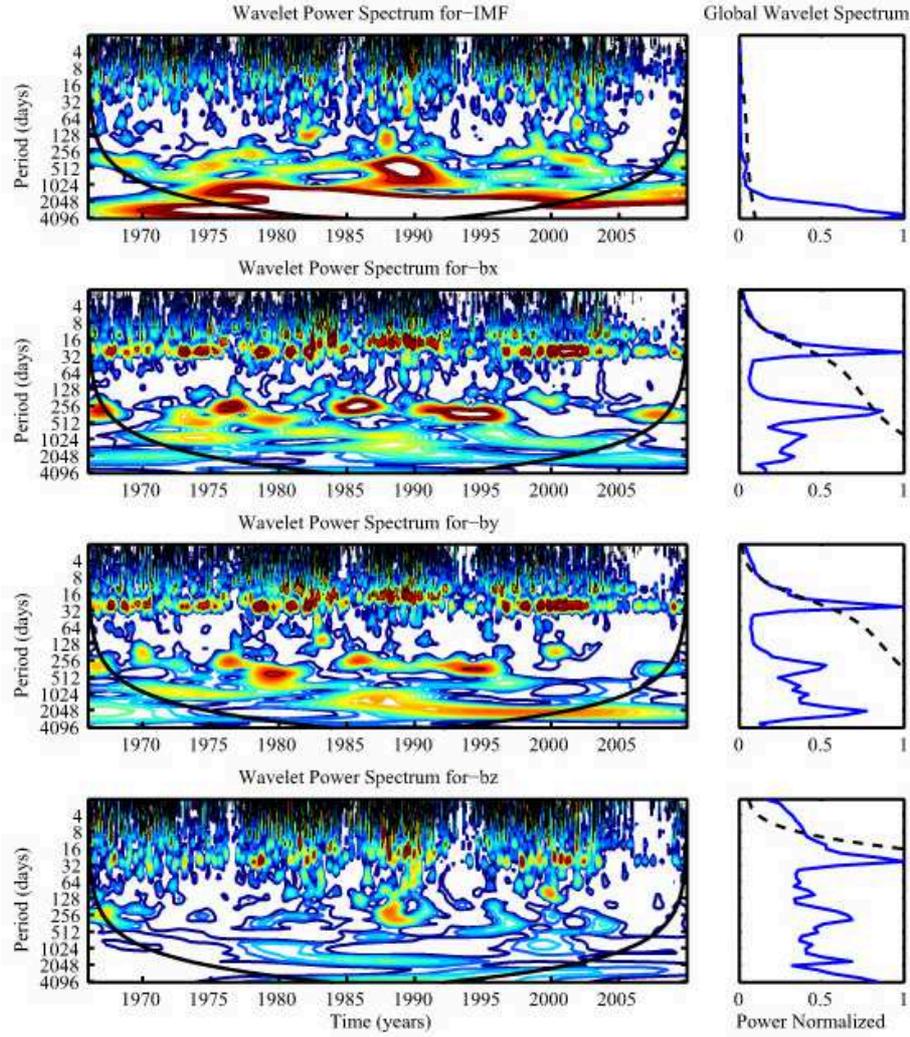}}
\caption{Same as Figure \ref{F1} for the interplanetary magnetic field (IMF) 
and its components B$_x$, B$_y$ and B$_z$. }
\label{F2}
\end{figure}
%-------------------------------------------------------------------------------------
\begin{figure}[ht]
\centering 
\resizebox{\textwidth}{!}{\includegraphics{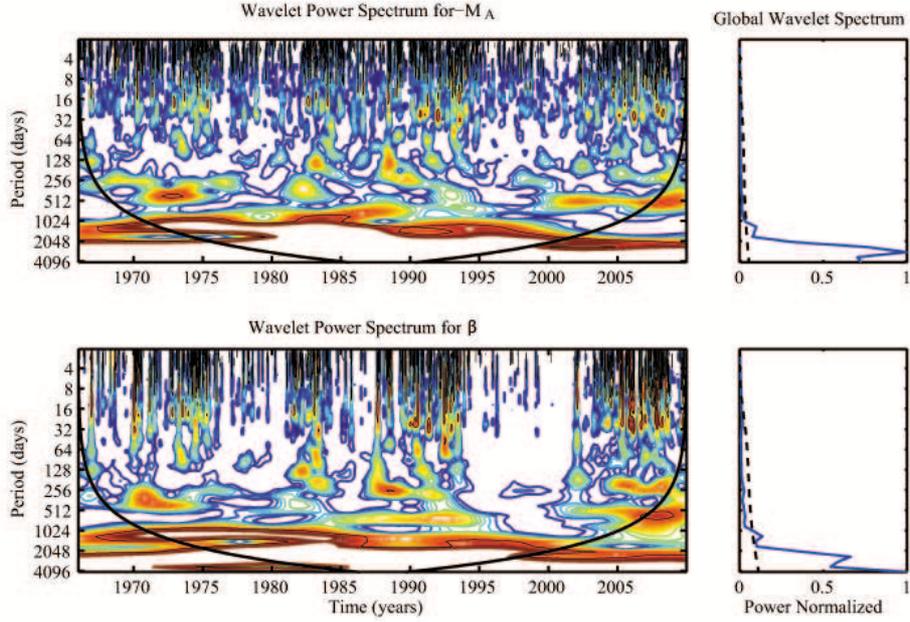}}
\caption{Same as Figures \ref{F1} and \ref{F2} Alfven Mach number ($M_A$, Upper panel) and plasma beta ($\beta$, lower Panel).}
\label{F3}
\end{figure}
%-------------------------------------------------------------------------------------
\begin{figure}[ht]
\centering 
\resizebox{\textwidth}{!}{\includegraphics{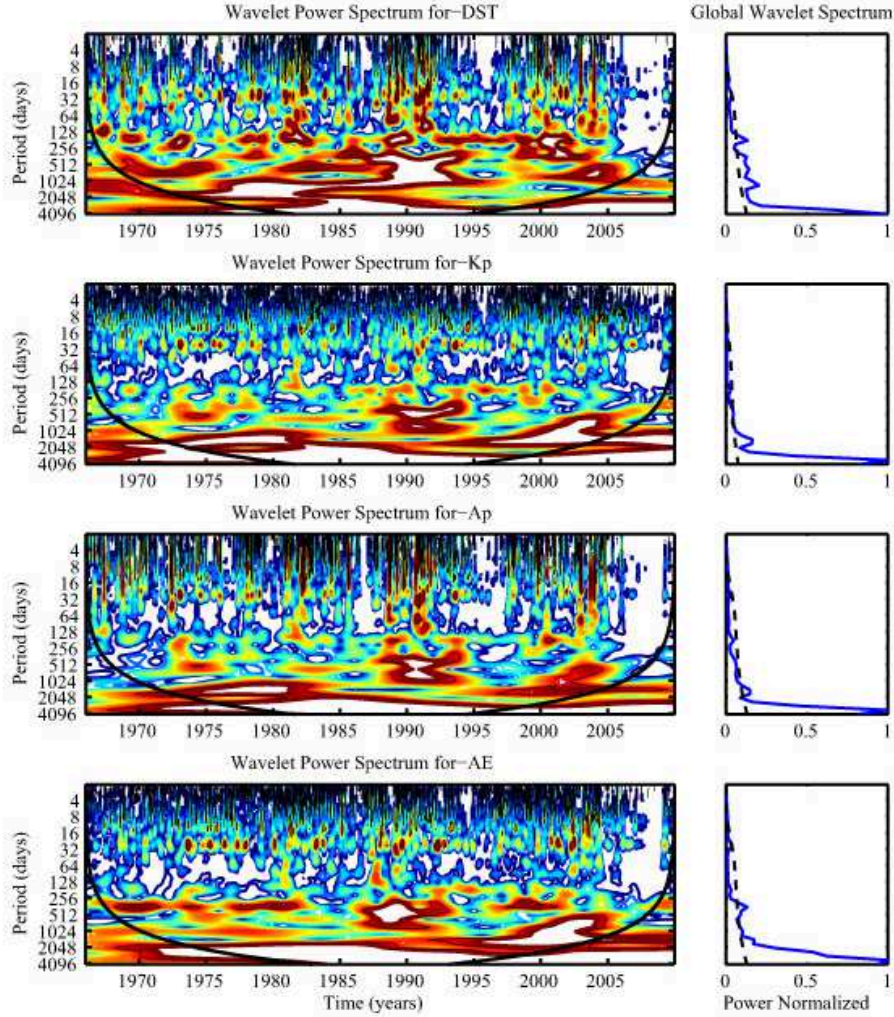}}
\caption{Same as Figures \ref{F1}, \ref{F2} and \ref{F3} for  the 
geomagnetic indices (Top to bottom)(D$_{ST}$, K$_{p}$, A$_{p}$ and AE). }
\label{F4}
\end{figure}
%-------------------------------------------------------------------------------------
\begin{table}[h!]
\caption{Periodicities (in days) indicated by Wavelet analysis. {\rd These periodicities span a range of
periods; the corresponding peak is within parenthesis next to the range. The b--superscript marks periodicities 
which are below confidence level of 99\% in the global wavelet spectrum. The column labeled \emph{interval} presents the 
length of time (occasionally more than one) within which the range of Mid--term periodicitities was detected; as similar column was omitted from the 
long-term periodicities as in this case the \emph{interval} was always equal to 44 years.}}
\begin{tabular}{cllll}
\hline
 {\bf{Par.}}  		&{\bf{Short--term}}    			&{\bf{Mid--term}}    		& {\bf{Interval}}	&{\bf{Long--term}}        \\
\hline
				&9-14 (13.9)      			&   256-512~(314)		&  7--8 years          	&                         \\
 V$_{SW}$     			&15-25$^{b}$         			&    512-1024~(629)		&  4--11 years         	&1024-4096~(1499, 2998)   \\
				&22-30 (27.8)     			&                      		&                  	&                          \\
\hline
				&9-22 (13.9)      			& 256--445$^{b}$~(314) 		&  3--4 years                 	&                         \\
 Temp.  			&15-25$^{b}$         			& 512--1024$^{b}$~(889)		&  6 years                 	&1024--4096~(1499, 2521)  \\
				&22-30$^{b}$(27.8)   			& 				&                   	&                         \\
\hline
 Density      			& 9-14$^{b}$         			&256--445$^{b}$~(314)  		& 3--4 years                  	&1024--4096~(1499, 2998)  \\
				& 15-25$^{b}$        			& 512--1024$^{b}$		&  10 years                 	&                         \\
\hline
Pressure      			&                 			&256--630~(528)			& 4 years                  	&1024--4096~(2998)        \\
\hline
IMF B         			&                 			& 256--1024~(444)    		&  6 years                 	&1024--4096~(3565)        \\
\hline
 B$_{x}$      			& 9-22~(13.9)          			& 256--512~(314)   		&  3--6 years                 	&                         \\
				& 22-30~(27.8)         			&  	     			&                   	&                         \\
\hline
 B$_{y}$      			& 9-22~(13.9)          			& 256--512$^{b}$~(314) 		&  3--4years                 	&                         \\
				& 22-30~(27.8)         			&  	        		&                   	&                         \\
\hline
B$_{z}$ 			& 9-22$^{b}$~13.9)    			&               		&                   	&                         \\
				& 22-30$^{b}$~(27.8)  			&               		&                    	&                   \\
\hline
M$_{A}$       			&                 			&128--512$^{b}$   		&  4--10 years                 	&1024--4096~(1260, 2998)  \\
				&                 			&               		&                   	&                         \\
\hline
$\beta$       			& 14--30$^{b}$~(27.8)			&128--512$^{b}$   		&  5 years                 	&1024--4096~(1260, 2521)  \\
				&       	       			&               		&                   	&                         \\
\hline
				&9--22 $^{b}$~(13.9)       		& 128--256~(187)		&  2--11 years                 	& 1024--4096~(1260, 4096) \\
D$_{ST}$      			&22--30~(27.8) 				& 256--512~(374)		&  6--15 years                 	&                         \\
				&64--128         			& 512--1024~(629)		&  7--15 years                 	&                         \\
\hline
  K$_{p}$     			&22-30$^{b}$~(27.8) 			& 128--256$^{b}$~(187)  	&  1--5 years                 	&1024--4096~(1499, 3565)  \\
				&  		    			& 256--1024~(374)		&   4--6 years                	&                         \\
\hline
  A$_{p}$     			&12-30$^{b}$~(13.9, 27.8)		&128--256$^{b}$~(187)		&   1--2 years                	&1024--4096~(1260, 3565)  \\
				&64--128$^{b}$      			& 256--1024$^{b}$~(374) 	&   3--6 years                	&                         \\
\hline 
  AE          			&22-30$^{b}$~ (27.8) 			&256--1024$^{b}$~(374) 		&   5--8 years                	& 1024--4096~(1499, 3565) \\
				&                    			& 	        		&                   	&                         \\
\hline
\end{tabular}
\label{table1}
\end{table}
%-------------------------------------------------------------------------------------
\begin{figure}[ht]
\centering 
\resizebox{\textwidth}{!}{\includegraphics{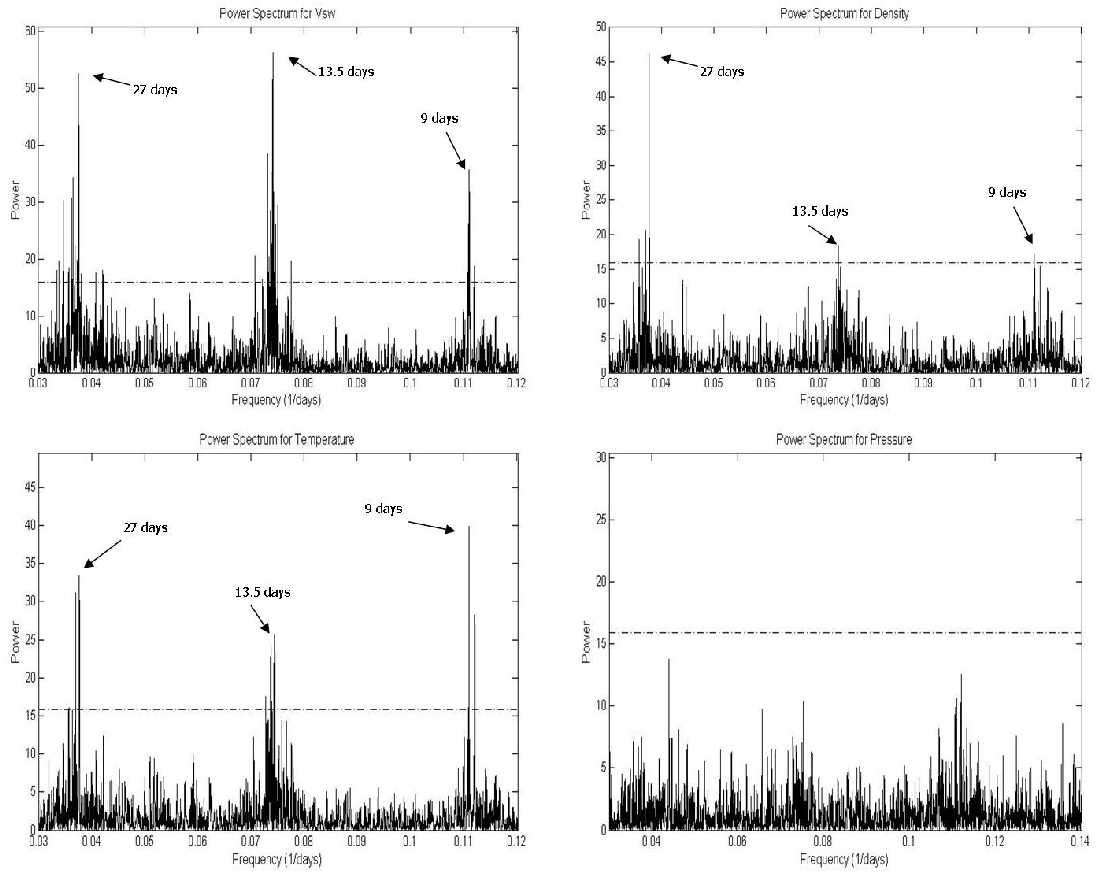}}
\caption{Lomb/Scargle periodogram for the short--term periodicities of the four dynamic 
parameters of the solar plasma: solar wind speed (Vsw), plasma temperature (T), plasma density (D) and plasma pressure (P). 
(The dashed line in all plots denotes the 99\% confidence level - peaks below this line are considered insignificant.)}
\label{F5}
\end{figure}
%-------------------------------------------------------------------------------------
\begin{figure}[ht]
\centering 
\resizebox{\textwidth}{!}{\includegraphics{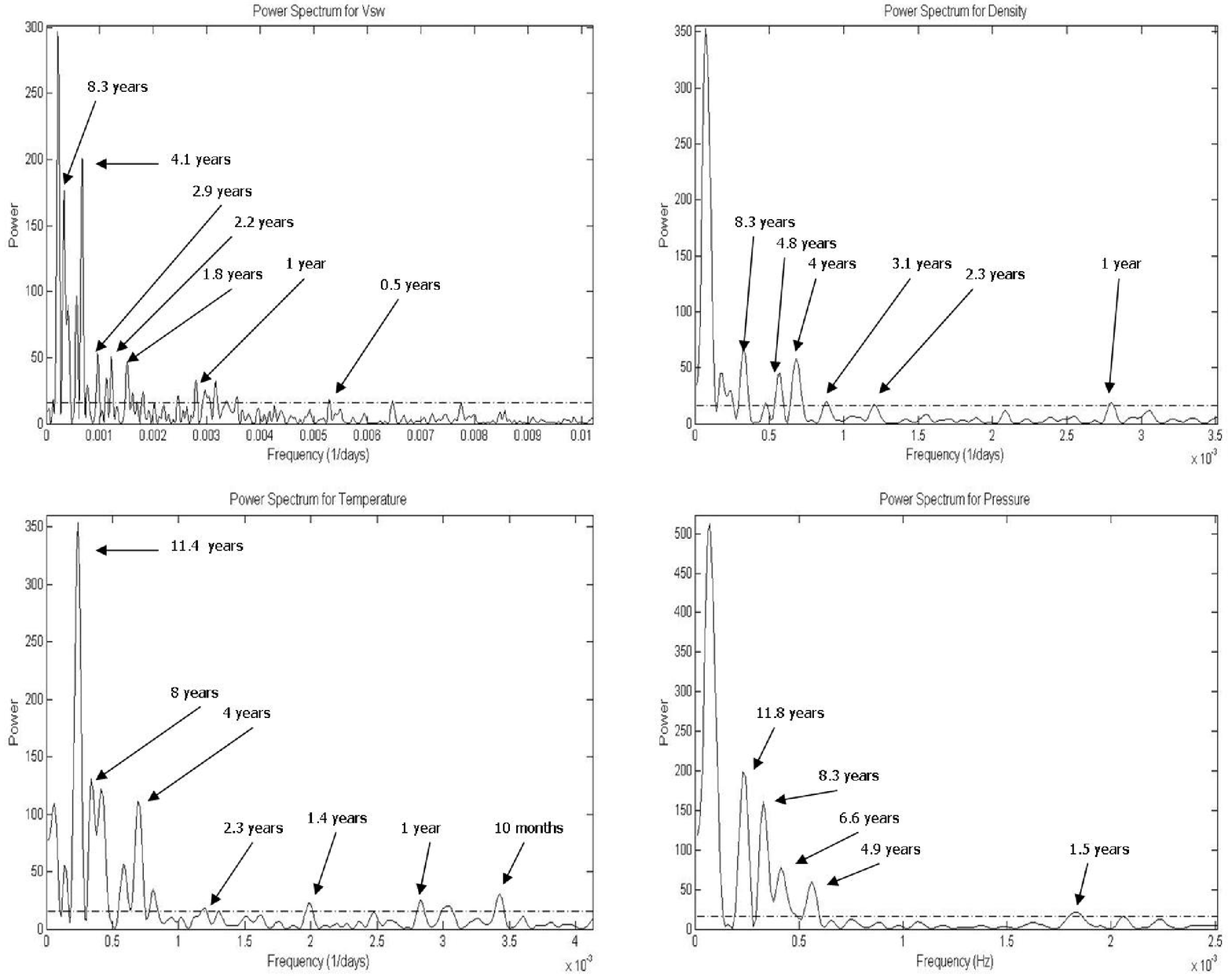}}
\caption{Lomb/Scargle periodogram for the long--term periodicities of the four dynamic parameters of the solar plasma: 
solar wind speed (V$_{sw}$), plasma temperature (T), plasma density (D) and plasma pressure (P). 
(The dashed line in all plots denotes the 99\% confidence level - peaks below this line are considered insignificant.)}
\label{F6}
\end{figure}
%-------------------------------------------------------------------------------------
\begin{figure}[ht]
\centering 
\resizebox{\textwidth}{!}{\includegraphics{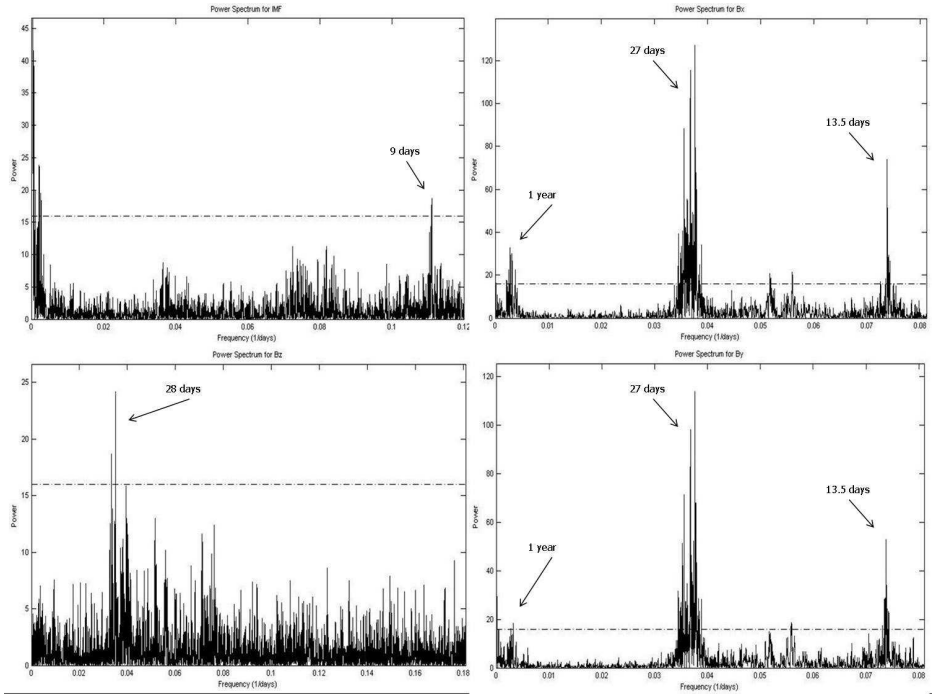}}
\caption{Lomb/Scargle periodogram of the interplanetary magnetic field (IMF) and its components x, y, z. 
(The dashed line in all plots denotes the 99\% confidence level - peaks below this line are considered insignificant.) }
\label{F7}
\end{figure}
%-------------------------------------------------------------------------------------
\begin{figure}[ht]
\centering 
\resizebox{\textwidth}{!}{\includegraphics{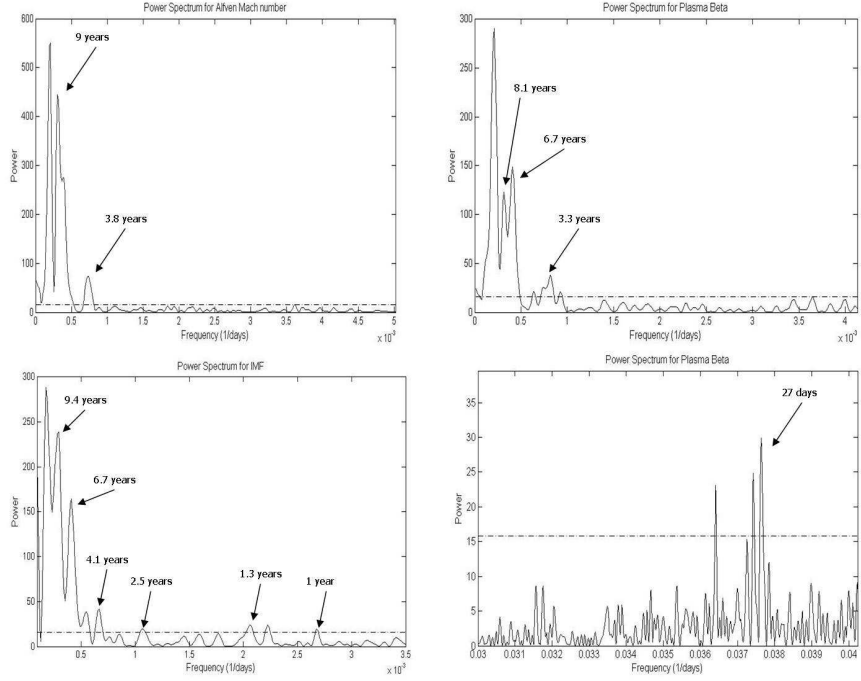}}
\caption{Lomb/Scargle periodogram for the long--term periodicities of the interplanetary 
magnetic field (IMF), Alfven Mach number and plasma Beta and the short--term periodicities of Plasma Beta. 
(The dashed line in all plots denotes the 99\% confidence level - peaks below this line are considered insignificant.)}
\label{F8}
\end{figure}
%-------------------------------------------------------------------------------------
\begin{figure}[ht]
\centering 
\resizebox{\textwidth}{!}{\includegraphics{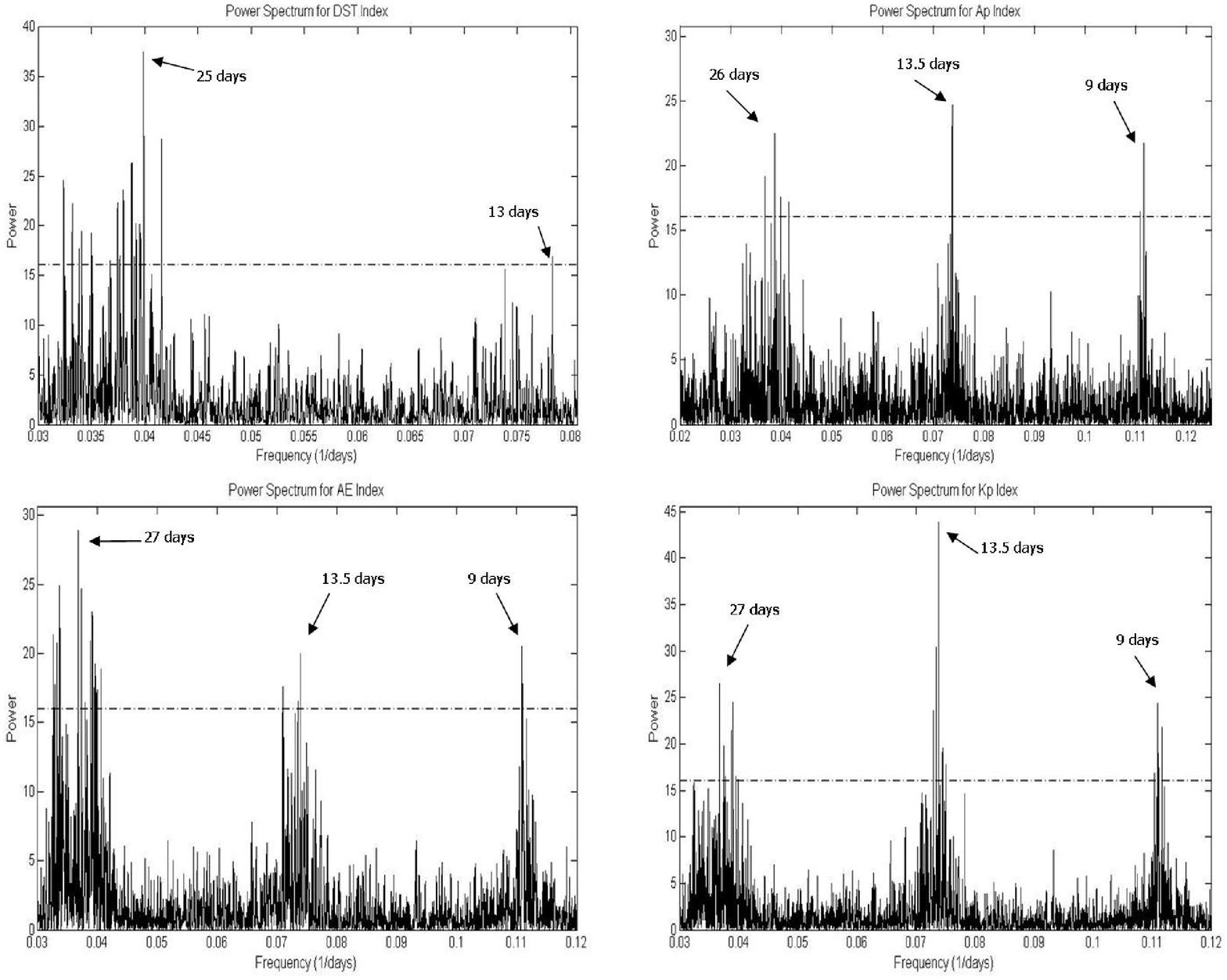}}
\caption{Lomb/Scargle periodogram for the short--term periodicities of the geomagnetic 
indices (D$_{ST}$, AE, K$_{p}$ and A$_{p}$ ). (The dashed line in all plots denotes the 99\% confidence level - peaks below this line are considered insignificant.}
\label{F9}
\end{figure}
%-------------------------------------------------------------------------------------
\begin{figure}[ht]
\centering 
\resizebox{\textwidth}{!}{\includegraphics{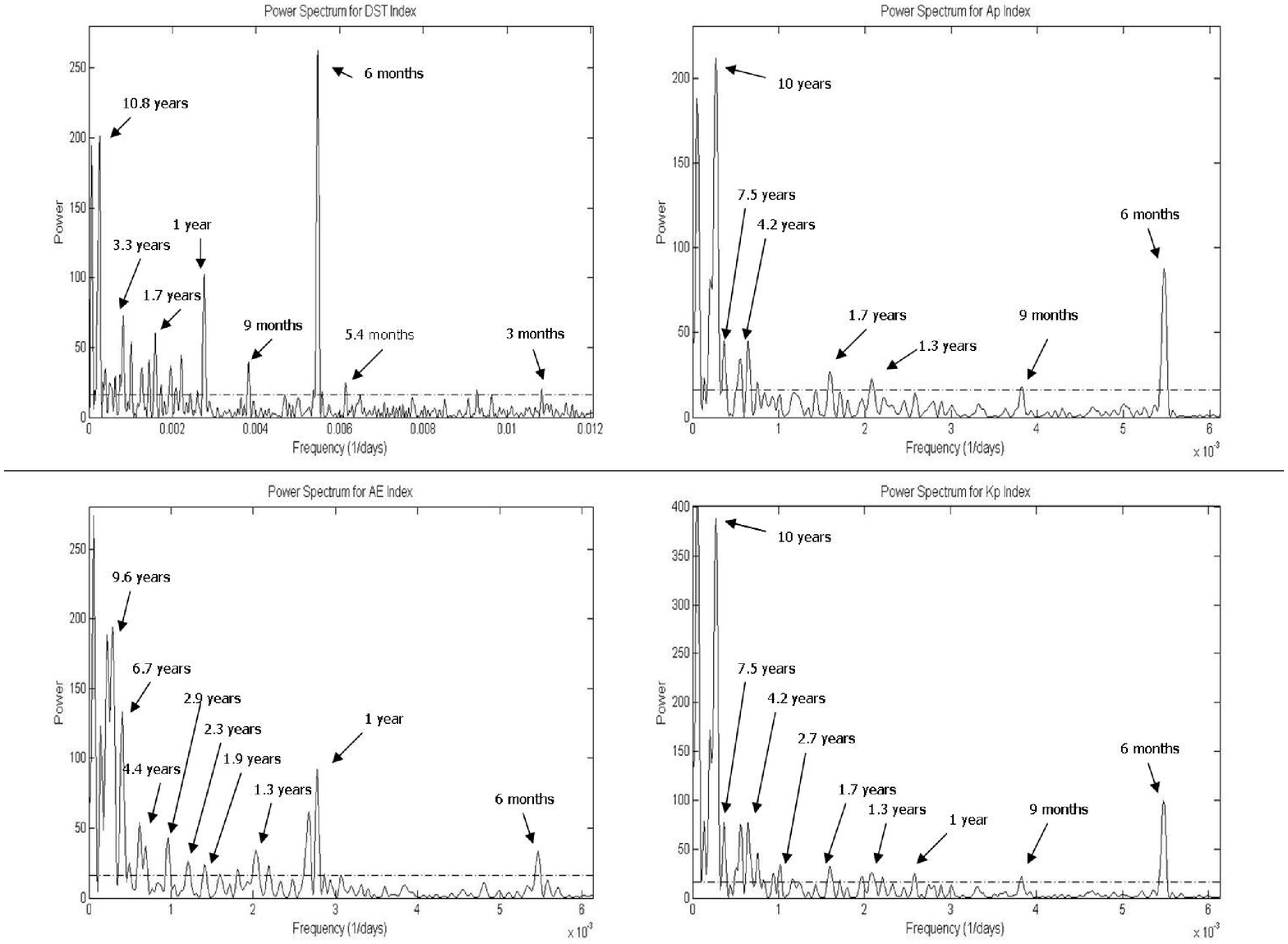}}
\caption{Lomb/Scargle periodogram for the long--term periodicities of the 
geomagnetic indices (D$_{ST}$, AE, K$_{p}$ and A$_{p}$ ). 
(The dashed line in all plots denotes the 99\% confidence level - peaks below this line are considered insignificant.)}
\label{F10}
\end{figure}
%-------------------------------------------------------------------------------------
\begin{table}[h!]
\caption{Table of periodicities (in days) revealed by the Lomb/Scargle periodogram. The confidence level is above 99\%. }
\begin{tabular}{cllll}
\hline
{\bf{Parameter}}  		&{\bf{Short--term}}    		&{\bf{Mid--term}}    		&{\bf{Long--term}}\\
\hline
V$_{SW}$     	& 9,~13.5,~27     		& 182,~316,~358,~554,~657,~823   	& 1032,~1513,~3026      \\
 Temperature 	& 9,~13.5,~27     		& 291,~329,~502,~353,~834		&  1441,~2937,~4173   \\
 Density     	&9,~13.5,~27     		&  357,~830    				& 1126,~1468,~1749,~3045   \\
 Pressure    	&             			&  316,~357,~544         		& 1788,~2403,~3045,~4383       \\
\hline
 IMF B       	& 9           			&  373,~485,~934         		&  1507,~1827,~2458,~3446  \\
 B$_{x}$     	&13.5~27       			& 353          				&          \\
 B$_{y}$     	&13.5~27        		& 313,~353          			&          \\
B$_{z}$      	&28~30          		&              				&          \\
M$_{A}$      	&             			&              				&    1376,~3268    \\
$\beta$      	&  27         			&              				& 1220,~2459,~2956    \\
\hline
D$_{ST}$     	& 13,~25       			& 182,~261     				&  1204       \\
		& 92          			& 362,`632          			&  3942      \\
\hline  
 		& 9           			& 182,~262          			&  1541    \\
  K$_{p}$    	& 13.5        			& 388,~81      				&   2739   \\
		& 27          			& 986,~624          			&  3652   \\
\hline
		& 9           			&182           				&   1541    \\
  A$_{p}$    	& 13.5        			& 261          				&   2739     \\
		& 26          			& 481,~628     				&   3652   \\
\hline
		& 9           			& 183,~360          			& 1055,~1601   \\
  AE         	& 13.5        			& 492,~711     				&  2441    \\
		& 27          			& 828          				&  3503  \\
\hline
\end{tabular}
\label{table2}
\end{table}
%-----------------------------------------------------------------------------------------------
\section{Observations and Data Analysis} \label{Event}
\subsection{Data Selection}\label{DSel}
The solar wind parameters examined are speed ($V_{sw}$), proton density (D), temperature (T), plasma pressure (P), 
inter-planetary magnetic field (B$_{x}$, B$_{y}$, and B$_{z}$), Alfven Mach number ($M_A$, local flow velocity to the local Alfven speed), 
and plasma beta ($\beta= nkT/[B^2/{\mu}_0]$, plasma pressure to the magnetic pressure ratio). The corresponding geomagnetic indices 
studied are, the $D_{ST}$ 
{\rd{(a measure of the strength of Earth's ring current)\footnote{\rd The $D_{ST}$ Index is 
calculated by averaging the horizontal component of the magnetic field from mid--latitude and 
equatorial magnetograms}}}, 
the $K_p$ (mean value of the disturbance levels in 
the two geomagnetic horizontal field components), the $A_p$ (linear version of $K_p$) and Aurora Electrojet index AE. The latter 
is a global, quantitative measure of auroral zone magnetic activity produced by enhanced ionospheric currents flowing 
below and within the auroral oval. 

The OMNI database\footnote{http://omniweb.gsfc.nasa.gov} was used and all data are measured at the 
distance of 1 A.U. in daily values from January 1st, 1966 to December 31st, 2010 (last four solar cycles).  

\subsection{Wavelet analysis}\label{WAnal}

{\rd As we analyze non--stationary time--series we expand 
in terms of time--localized waves, or {\em{wavelets}} in order to obtain a compact, two dimentional, 
representation \citep[see~][]{Morlet1982,Torrence1998} of these time--series: 

\[y\left( {t,t',f} \right) = \exp \left( {2i\pi ft} \right)\exp \left( { - f^2 \frac{{\left( {t - t'} \right)^2 }}{2}} \right)\]
where f is the frequency, t' the time delay and \[exp \left( { - f^2 \frac{{\left( {t - t'} \right)^2 }}{2}} \right)\] is the Gaussian support.

We use the {\em{Morlet wavelet}} because it's the most common wavelet function used for revealing periodicities 
of astrophysical signals; this makes easier the comparison of our results with previously published works. Furthermore, 
due to its Gaussian support, the Morlet {\em{wavelet}} expansion inherits optimality as regards the uncertainty principle \citep{Morlet1982}.

We also calculate the global wavelet spectrum \citep[see~][]{Torrence1998} which 
is the time-averaged wavelet spectrum 
when the average is over all the local wavelet spectra and is given by the equation
\be
\overline{W}^{2}(s) = 1/N \sum^{N-1}_{n = 0} \left|W_{n}(s)\right|^{2}
\ee
where $W_{n}(s)$ is the wavelet power and N the number of local wavelet spectra. 
Thus we obtain an unbiased and consistent estimation of the true power spectrum of a time series. }

\subsection{Wavelet spectra of solar wind parameters and geomagnetic indices}\label{WSpec}

In this section we examine the variation of the parameters of the solar wind and the geomagnetic indices; these provide us with 
fourteen time series in total. We use the wavelet power spectra and the global wavelet spectra for the identification of periodic 
components to a confidence level\footnote{\rd The confidence interval is defined as the probability that the true wavelet power at a certain time and
scale lies within a certain interval about the estimated wavelet power.} of 99\%.

Starting from the periodicities of the solar wind speed, temperature, density and pressure (figure \ref{F1}) we note certain 
similarities as follows:

\begin{itemize}
\item{ The long--term periodicities (1024--4096 days) appear throughout the observation period regarding speed, temperature, 
density (with peaks at 4.1 and 8.2 years) and pressure (peak at 8.2 years only).}

\item Mid--term periodicities (256--512 days with peak at 314 days) in speed appear in 1970--1977 and 1985--1992. These appear also 
in temperature and density (peak 314 days). Periodicities of 512--1024 
days appear in speed (peak 1.7 years) at the rising phase and maximum of cycle 20 and at cycles 22 and 23 and (below confidence 
level) in temperature (peak 2.4 years) at cycles 22 and 23 and pressure (peak 528 days).

\item{Short--term periodicities of 9--14 (peak 13.9) and 22--30 (peak 27.8) days, corresponding to the rotation of the sun and 
sub-harmonic, appear quite pronounced in the speed during the descending phases of cycles 20 and 23 but less pronounced during 
cycles 21, 22. These short-term periodicities appear in temperature and density as well but below the 99\% confidence level. }

\end{itemize}

Next, we isolate the periodic components of the interplanetary 
magnetic field examining B, B$_{x}$, B$_{y}$, B$_{z}$, Alfv\'{e}n Mach number 
and plasma beta (Figures \ref{F2} and \ref{F3}): 

\begin{itemize}

\item{ Long--term periodicities of (1024--4096 days) days appear throughout the observation period in IMF B (peak 9.8 years), 
Alfv\'{e}n Mach number (peaks 3.5 and 8.2 years) and plasma beta (peaks 3.5 and 6.9 years).}

\item{Mid--term periodicities (256--1024 days with peak at 1.2 years) appear around the maximum of cycle 22 in IMF B. A range of 256--512 days (peak 314 days) 
periodicities appear in B$_{x}$ (and B$_{y}$ below the confidence level) during the ascending phases of cycles 20, 21 and 22 and the descending phase of 22. 
The mid--term periodicities in Alfv\'{e}n Mach number and plasma beta are below the confidence level.}

\item{Short--term periodicities of 9--22 days (peak 13.9 days) and 22--30 days (peak 27.8 days) appear in B$_x$ and B$_y$ (B$_z$ and plasma beta below the confidence level) throughout the observation period except the minima of cycle 22 and 23 around 1995 and 2005.}

\end{itemize}

Finally we examine the periodicities of geomagnetic indices D$_{ST}$, K$_{p}$, A$_{p}$ and AE (Figure \ref{F4}):

\begin{itemize}

\item{Long--term periodicities of 1024--4096 days appear throughout the whole observation period in 
D$_{ST}$ (peaks 3.5 and 10.8 years), K$_{p}$ (peaks 4.1 and 9.8 years), A$_{p}$ (peaks 3.5 and 9.8 years) and AE (peaks 4.1 and 9.8 years).}

\item{Mid--term periodicities in geomagnetic indices vary: A periodicity range of 256--1024 days (peak 1 year) in AE, K$_{p}$ and A$_{p}$ (below confidence level) appears during the maximum and descending phase of cycle 22. The same extends in D$_{ST}$ (with peaks 1 year and 1.7 years) 
		but in the time intervals 1969--1975, 1980-1995 and 2000--2005.

	\begin{itemize} 

	\item{A smaller range of periodicities of 512 -- 1024 days appear in K$_{p}$, A$_{p}$ and AE in 2000--2005.}

	\item{A smaller range of periodicities of 256--512 days (peak 1 year) appear around the maximum of every cycle in AE.}

	\end{itemize}}

\item{ Short--term periodicities (22--30 days, with 27.8 days peak) appear intermittently and without any obvious pattern in all four indices (above the confidence level only D$_{ST}$). 
Periodicities of 64--128 days are observed in D$_{ST}$ (maxima of cycles 21, 22, 23 and descending phase of 23) and A$_{p}$ (maximum of cycle 22 
and descending phase of 23 below the confidence level in both cases). Also periodicities of 128--256 days (peak 187 days) appear in D$_{ST}$ throughout 
the observation period up to 2004, these periodicities extend 
to K$_{p}$ and A$_{p}$ but below the confidence level.}

\end{itemize}

\subsection{Lomb/Scargle method}\label{Lomb}

\citet{Scargle1982} modified the standard periodogram formula to first find a time delay $\tau$ such 
that the power of each frequency was indepedent of any constant shift of sample time $t_{j}$ and equivalent to the reduction of the sum
of squares in least-squares fitting of sine waves to the data. The time delay $\tau$ is defined as:
\be
tan~(2\omega\tau) = \frac{\sum_{j} sin (2\omega t_{j})}{\sum_{j} cos (2\omega t_{j})}
\ee     
The periodogram at frequency $\omega$ is then estimated as:
\be
P_{x} (\omega)=1/2 \frac{[\sum_{j} X_{j}cosù (t_{j}-\tau)]^{2}}{\sum_{j} cos2ù(t_{j}-\tau)}+1/2 \frac{[\sum_{j} X_{j}sinù(t_{j}-\tau)]^{2}}{\sum_{j} sin2ù(t_{j}-\tau)}
\ee
{\rd where $\omega$ is the frequency and X$_{j}$ is the value of the physical quantity measured in time $t_{j}$ ,
The Lomb/Scargle periodogram, being a variant of the Fourier transform with application in unevently spaced data, 
represents a signal as the sum of sine and cosine functions of infinite duration. More often than not, however,
the signal statistical properties, characteristic periodicities included, vary over time so 
this method provides information about the frequency content of a time series without 
localizing it in time. The discrete periodicities in the Lomb/Scargle periodogram sometimes correspond to a range of periodicities 
in the wavelet spectrum.}

\subsection{Lomb/Scargle periodogram of solar wind parameters and geomagnetic indices}\label{LombAnal}

In this section we present the Lomb/Scargle periodogram for each parameter of the solar wind and geomagnetic 
index (figures \ref{F5} to \ref{F10}). The periodicities we accept are those above the confidence level of 99 \%.

In figure \ref{F5} and \ref{F6} we present periodicities of the four parameters of the solar wind (speed, 
temperature, density and pressure). There are similarities at speed, density and temperature as stated also in section \ref{WAnal}: 

\begin {itemize}

\item {The short--term periodicities of 9, 13.5 and 27 days appear in all three parameters (figure \ref{F5}).}

\item {The annual periodicity appears in all four parameters (figure \ref{F6}) but with slight 
differences (316 and 358 days in speed, 329 and 353 in temperature, 357 in density and 316 and 357 in pressure). 
There is also a semi--annual periodicity only in speed.}

\item {There are periodicities of 554 ad 657 days in speed, 502 days in temperature and 544 days in pressure. }

\item {A 2.2--2.3 years periodicity (823, 834 and 830 days) in speed, temperature and density respectively is noted. }

\item {Long--term periodicities appear in all four parameters with maxima at 2.9, 4.1 and 8.3 years for speed, 4, 8 and 11.4 years for 
temperature, 3.1, 4, 4.8 and 8.3 years for density and 4.9, 6.6, 8.3 and 11.8 years for pressure.}
\end{itemize}

In figures \ref{F7} and \ref{F8} we present the interplanetary magnetic field, Alfv\'{e}n Mach number and plasma beta. 

\begin {itemize}

\item {There is an identical behaviour of the x and y component of the IMF where we have the appearance of three periodicities of 13.5 and 27 days and 1 year. }

\item {IMF appears a periodicity of 9 days (figure 7) and mainly long--term periodicities of 1, 1.3, 2.5, 4.1, 5, 6.7 and 9.4 years (figure \ref{F8}). }

\item {The z component seems to appear only a 29--30 days periodicity.}

\item {Long term periodicities are mainly appearing in Alfv\'{e}n Mach number (figure \ref{F8}) with peaks of 3.8 and 9 years. }

\item {Plasma beta appears a periodicity of 3.3, 6.7 and 8.1 years and also the 27 days periodicity.}

\end{itemize}

The Lomb/Scargle periodograms in geomagnetic indices are presented in figures \ref{F9} and \ref{F10}.

\begin {itemize}

\item {There are once again similarities in K$_p$, A$_p$ index. Short--term periodicities (figure \ref{F9}) 
of 9, 13.5 and 27 days appear in K$_p$, A$_p$ and AE, 
while D$_{ST}$ shows periodicities of 13 days and 25 days. }

\item {Periodicities of 0.5 and 0.7 years (figure \ref{F10}) appear in K$_p$ and A$_p$ index as well as in D$_{ST}$, while AE contains only the 6 month periodicity. 
The D$_{ST}$ shows also a 92 days periodicity. }

\item {Long--term periodicities of 1--2.8 years appear in all indices but with different power and maxima. 
Maxima of 1.3, 1.7, 4.2, 7.5 and 10 years in A$_{p}$ and 1.3, 1.7, 2.7, 4.2 7.5 and 10 in K$_{p}$ are the most important, while AE 
index shows peaks of 1, 1.3, 1.9, 2.3, 2.9, 4.4, 6.7 and 9.6 years.}

\item {D$_{ST}$ shows 1, 1.7, 3.3 and 10.8 years maxima.}

\end{itemize}

\section{Summary and discussion}\label{SumDisc}

A number of short and long--term periodicities were detected within the 99\% confidence level.
We, firstly, draw attention to similarities in periodic behaviour between certain parameters of the solar wind and the magnetosphere. 

\begin{enumerate}

\item {Solar wind speed and temperature are expected to be correlated  \citep[see the study~][~for cycle 23]{Katsavrias2010}; 
they share the same periodicities in our data. }

\item {There is symmetrical behavior of the x and y component (B$_x$, B$_y$) of the IMF, so they exhibit identical periodicities}

\item {As $\beta\approx M_{A}^2$ they, too, share the same periodicities.}

\item {Similar behaviour is also expected from K$_p$ and A$_p$ index since the latter is a linear version of the former.}

\end{enumerate}

\subsection{\rd Short--term periodicities} 
{\rd These short--term periodicities, due to the solar rotation,} were detected in solar wind (V$_{SW}$) speed with enhanced amplitude during the descending phases of cycles 20 and 23 but 
less pronounced during cycles 21, 22; the latter cycles had larger maxima and shorter minima. Short--term periodicities appeared in temperature and density as well (but below the accepted 99\% confidence level) outside peaks of cycles. 
\citet{Nayar2002} observed these periodicities of V$_{SW}$ in the descending phases of 21, 22 and 23, while \citet{Bolzan2005}, 
report similar findings in speed and density in the years 1997 (less pronounced) and 2000. 
Periodicities of about 27.5, 13.5, 9.1, and 6.8 days were identified in the solar wind speed for the time interval 1964--2000 \citep{Gonzalez1987, Gonzalez1993, Svalgaard1975, Fenimore1978}. We found the 9, 13.5 and 27 days periodicities using the Lomb/Scargle periodogram. It's been proposed that intervals of large 13.5-day periodicity are due to the occurrence at 1 AU of two high-speed streams per solar rotation \citep{Mursula1996}.

{\rd In V$_{SW}$ (and T below the 99\% confidence level) the 27 days periodicity is quite apparent during all four solar cycles, covered in our date set, apart from each cycle's maximum. This is probably due to the fact that fast streams from coronal holes overlaps with mass ejection from or near the active regions resulting in a, more or less, random variation of the measured solar wind speed. In support of this line of reasoning, we note, that in cycles 21 and 22 where the maxima have benn quite pronounced and the minima of short duration, this effect of the suppression of the spectral peak in V$_{SW}$ was more evident compared to cycles 20 and 23.}

Periodicities of 27.5, 13.5, 9.1, and 6.8 days were identified in the IMF polarity for the time intervals 1964--2000 \citep{Gonzalez1987,Gonzalez1993,
Svalgaard1975,Fenimore1978,Mursula1996,Nayar2002}. We found short--term periodicities of 9--22 days (peak 13.9 days) and 22--30 days (peak 27.8 days) appear in Bx throughout the observation period except the descending phases of cycle 22 and 23. These appear also in By and Bz but below confidence level. Lomb/Scargle periodogram showed the same periodicities.

\subsection{Annual and semi--annual periodicity}
CMEs show an annual and semi--annual periodicity during the time interval 1994--2000 \citep{Polygiannakis2002} and an annual, semi--annual and 3 years periodicity during 1999--2003 \citep{Lou2003}. We found the annual periodicity in the Lomb/Scargle periodogram of speed, density, temperature, pressure, IMF, B$_x$ and B$_y$. Moreover, a semi-annual periodicity was found in speed and a 3 years periodicity in speed and density. This could be a strong argument for the dependence of the solar wind speed on CMEs but wavelet spectrum of solar wind show these periodicities (0.5 and 1 year) during solar cycles 20, 21, 22 and not the 23rd. In addition, the annual periodicity appear with enhanced amplitude in the cycles 20 and 22 in speed, density and temperature.

{\rd According to \citet{Lou2003}, A$_p$ index shows also periodicities of 187, 273 and 364 days during the time interval 1999--2003. We found a 182 days 
variation in D$_{ST}$ throughout the whole observation period which in the other indices is below the confidence level. The 273 and 364 days periodicities are included in the range of 256--1024 days (peak 1 year) in K$_{p}$ and (below confidence level) A$_p$ during the maximum and descending phase of cycle 22 and in AE during the whole observation period except the minima. This indicates that CMEs are correlated with the geomagnetic storm disturbances via solar wind speed.} 
The Lomb/Scargle periodogram shows a peak of 182 days in all four indices and a peak of 262 days (0.7 years) in D$_{ST}$, K$_p$ and A$_p$.

\subsection{1--2 years intermittent periodicity}
{\rd According to \citet{Obridko2007} the IMF shows a 1.3 years periodicity . We found this periodicity but only around the maximum of the cycle 22 
which is probably been propagating from the solar magnetic field (SMF), with no obvious explanation about its absence from the other cycles.} 

{\rd As reported by \citet{Gazis1995} and \citet{VGalicia1996} solar wind speed, CMEs and galactic cosmic rays show intermittent and somewhat varying (1--1.3 years) periodicity during 1974--2000 and 1964--2000 respectively.  \citet{Kudela2010} report that a range of periodicities 1.7--2.2 years, appear in Cosmic Rays during the time interval 1951--2010 and \citet{Mavromichalaki2003} during 1953–-1996. The same periodic modulation has been detected when fluxes of solar energetic protons and galactic cosmic rays are investigated \citet{Laurenza2009}. We found periodicities of 512--1024 days in speed (peak 1.7 years) at the ascending phase and maximum of cycle 20 and at cycles 22 and 23, as well as -- below the confidence level -- in temperature (maximum and descending phase of 22 and 23)and density (maximum and descending phase of 22). This is supported by the Lomb/Scargle periodogram that shows peaks of 1.5 and 1.8 years in speed and 1.4 years in temperature. It seems that there is a periodicity varing between 1.2--1.3 years and a periodicity of 2 years varing between 1.7--2.2 years.}

\citet{VGalicia1996,Mursula1999,Nayar2002} report different periodic variations of the geomagnetic activity index A$_p$; 1.3--1.4 years during {\em{even}} cycles and of 1.5--1.7 years during {\em{odd}}. 
{\rd The A$_p$ (and AE below the confidence level) periodicities measured were within the 1.3--1.7 years range yet no distinction between {\em{even}} or {\em{odd}} cycles was apparent.} 
On the other hand we found two intervals of periodic variations in all four indices (D$_{ST}$, K$_p$, AE and A$_p$), as follows:
 \begin{enumerate}
 \item {In the 256--1024 days range from 1990 to 1995; these coincide with variations, at the same frequency range within the same interval, of the V$_{SW}$, IMF (and T below the 99\% confidence level).}
 \item {512--1024 days in 2000--2005; these coincide with variations, at the same frequency range within the same interval, of the V$_{SW}$ (and T below the 99\% confidence level) but not IMF this time.}
 \end{enumerate}
We interprete this association as the result of the Solar Wind and IMF triggering and modulating magnetospheric activity. The effects seem to spread to all latitudes (K$_p$, AE) and to the ring current as well (D$_{ST}$).

The Lomb/Scargle periodogram confirmes the 1.7--1.8 years periodicity in V$_{SW}$, D$_{ST}$, K$_p$, A$_p$ and AE (below the 99\% confidence level); this corroborates similar results by 
\citet{VGalicia1996}. 

\subsection{Long--term periodicities}
Periodicities of 1024--4096 days appear throughout the whole observation period in solar wind speed (peaks 4.1 and 8.2 years), 
temperature (peaks 4.1 and 6.9 years), density (peaks 4.1 and 8.2 years) and pressure (peak 8.2 years). These also extend to D$_{ST}$ (peaks 3.5 and 11.2 years), 
K$_{p}$ (peaks 4.1 and 9.8 years), Ap (peaks 3.5 and 9.8 years) and AE (peaks 4.1 and 9.8 years). The corresponding Lomb/Scargle periodogram gives us also peaks 
of 2739 days (7.7 years) and 1204 days (3.3) in K$_p$ and A$_p$. The former was also found in cosmic rays in 1953–-1996 
\citep{Mavromichalaki2003} and the latter in 1947--1990 \citep{VGalicia1996} .

A 9.8 years periodicity was also found throughout the four cycles corroborating findings of \citet{Nayar2002} in 1964--2000. 
In the Lomb/Scargle periodogram appear also harmonics to the above mentioned periodicity of 6.7, 5, 4.1 and 2.5 years.  

\subsection{Conclusions}

{\rd Both the wavelet analysis and the Lomb/Scargle periodogram, have identified the 27 days periodicity (with 13.5 days being its harmonic) 
in the dynamic parameters of the solar wind, Bx, By and the geomagnetic indices. The 1--1.4 years range of periodicities  in 
the geomagnetic indices, the average IMF and  solar wind speed and temperature was also identified. Furthermore the quasi--biennial (1.7--2.2 years) 
oscillation, along with its harmonics of 4 and 8 years, in all solar wind parameters, apart from IMF, and geomagnetic indices were detected.
In solar cycle 22 the periodicities were more clearly defined than in the rest of the observation period, with well pronounced spectral peaks.}

%-------------------------------------------------------------------------------------------
\begin{acks}
This work was supported in part by the university of Athens research center (ELKE/EKPA). 
Christos Katsavrias thanks the LOC of the ESPM 13 for the financial support. 
The authors appreciate discussions with Alexander Hillaris. They also 
acknowledge many useful comments by the anonymous reviewer and the full attention of the editor.
\end{acks}
%-------------------------------------------------
%\bibliographystyle{spr-mp-sola}
%\bibliographystyle{spr-mp-sola-cnd}
%\bibliography{LaTeX1}

%-------------------------------------------------------------------------------------
\end {article}
\end {document}